\documentclass[aps,prl,reprint,longbibliography,superscriptaddress]{revtex4-1}
\usepackage{mathrsfs}
\usepackage{amsmath}
\usepackage{amsfonts}
\usepackage{amssymb}
\usepackage{amsthm}
\usepackage{graphicx}
\usepackage{natbib}
\usepackage{color}
\usepackage{hyperref}
\usepackage{bm}
\usepackage[caption=false]{subfig}
\usepackage{verbatim}

\newcommand{\mats}[1]{ \left \langle \mathrm{T}_{\tau}  #1 \right \rangle }

\begin{document}
	\title{Spin splitting induced in a superconductor by an antiferromagnetic insulator}
	\author{Akashdeep Kamra}
	\affiliation{Department of Physics, University of Konstanz, D-78457 Konstanz, Germany}
	\affiliation{Center for Quantum Spintronics, Department of Physics, Norwegian University of Science and Technology, NO-7491 Trondheim, Norway}
	\author{Ali Rezaei}
	\affiliation{Department of Physics, University of Konstanz, D-78457 Konstanz, Germany}
	\author{Wolfgang Belzig}
	\email{wolfgang.belzig@uni-konstanz.de}
	\affiliation{Department of Physics, University of Konstanz, D-78457 Konstanz, Germany}
	
	\begin{abstract}
		Inspired by recent feats in exchange coupling antiferromagnets to an adjacent material, we demonstrate the possibility of employing them for inducing spin splitting in a superconductor, thereby avoiding the detrimental, parasitic effects of ferromagnets employed to this end. We derive the Gor'kov equation for the matrix Green's function in the superconducting layer, considering a microscopic model for its disordered interface with a two-sublattice magnetic insulator. We find that an antiferromagnetic insulator with effectively uncompensated interface induces a large, disorder-resistant spin splitting in the adjacent superconductor. In addition, we find contributions to the self-energy stemming from the interfacial disorder. Within our model, these mimic impurity and spin-flip scattering, while another breaks the symmetries in particle-hole and spin spaces. The latter contribution, however, drops out in the quasi-classical approximation and thus, does not significantly affect the superconducting state.
	\end{abstract}

	\maketitle
	
	
	{\it Introduction.} $-$ Conventional Bardeen-Cooper-Schrieffer (BCS) superconductors~\cite{Bardeen1957} are incompatible with magnetic interactions as the latter tend to break the Cooper pairing~\cite{Cooper1956} between the opposite-spin electrons. Nevertheless, the so-called Pauli contribution, associated with energy splitting of the two spin states, leads to interesting new phenomena when the spin splitting is comparable to the `unperturbed' superconducting gap~\cite{Maki1964}. These include spatially inhomogeneous order parameter in an otherwise homogeneous superconductor~\cite{Fulde1964,Larkin1965}, gapless superconductivity~\cite{Abrikosov1961,Maki1969}, and a first-order phase transition between superconducting and normal states~\cite{Clogston1962,Chandrasekhar1962}, all of which have been experimentally observed~\cite{Saint1969,Buzdin2005}. Furthermore, hybrids incorporating such spin-split superconductors were recently predicted~\cite{Machon2013,Machon2014,Ozaeta2014}, and found~\cite{Kolenda2016,Kolenda2017}, to exhibit large thermoelectric effects. The spin splitting in the superconducting layer may be induced by a magnetic field or via exchange coupling to a magnetic layer~\cite{Kolenda2017,Strambini2017} and leads to intriguing transport properties reviewed in~\cite{Bergeret2018}.
	
	The success of `exchange biasing' a ferromagnet (FM) layer via its coupling to an adjacent antiferromagnet (AFM) has been instrumental in the contemporary memory technology~\cite{Nogues1999,Fert2008,Nogues2005}. A simplified picture of exchange biasing in FM/AFM bilayers requires the AFM interface to be uncompensated, i.e. possess finite surface magnetization~\cite{Nogues1999,Stamps2000,Nogues2005}. Several theoretical models~\cite{Stamps2000}, most of which assume the AFM surface to be uncompensated, have been employed to understand the experiments. Recent progress in surface characterization methods~\cite{Srajer2006} and epitaxial sample growth~\cite{Zhang2016} has enabled to resolve~\cite{Manna2014,Zhang2016} several previously open questions~\cite{Nogues1999}. Numerous experiments~\cite{Zhang2011,Kappenberger2003,Sampaio2003,Camarero2006,Roy2005,Valev2006,Ohldag2003,Blomqvist2004,Mathieu1998} have succeeded in direct observation and quantification of uncompensated spins at interfaces thereby improving the understanding of their role in exchange bias and the control of the effect.
	
	Recently, the presence of surface magnetization, stemming from broken translational symmetry at interfaces, in magnetoelectric AFMs has been predicted~\cite{Belashchenko2010}. This has also been observed experimentally and exploited in achieving electrically switchable exchange bias~\cite{He2010} and magnetic memory~\cite{Kosub2017} using $\alpha$-$\mathrm{Cr}_{2} \mathrm{O}_{3}$. Furthermore, uncompensated AFM interfaces have been theoretically predicted to amplify transfer of magnonic spin from a magnetic insulator to an adjacent non-magnetic conductor~\cite{Kamra2017,Bender2017}.

	\begin{figure*}[tbh]
		\begin{center}
			\subfloat[]{\includegraphics[width=35mm]{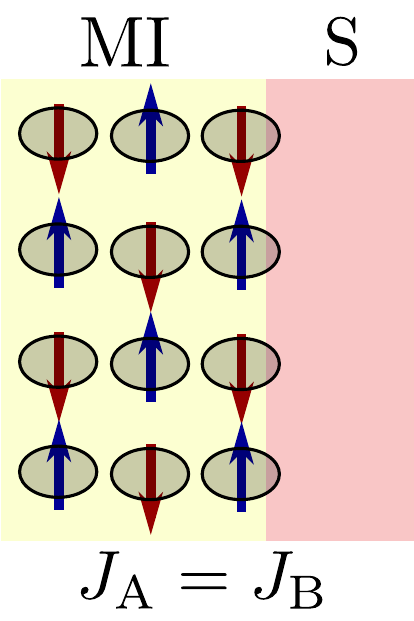}} \qquad
			\subfloat[]{\includegraphics[width=35mm]{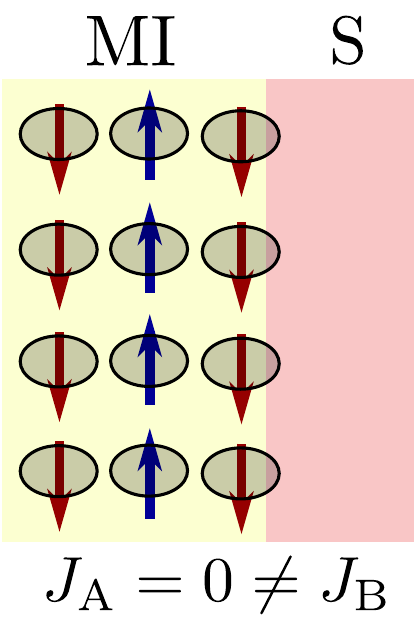}} \qquad
			\subfloat[]{\includegraphics[width=35mm]{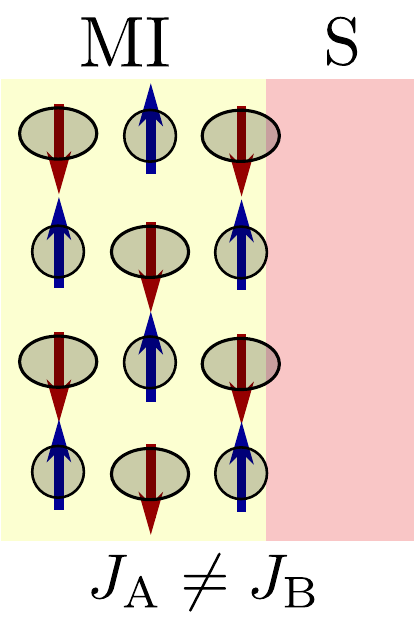}}
			\caption{Possible interface microstructures for magnetic insulator (MI)/superconductor (S) bilayers. Sublattices A and B are depicted in blue and red, respectively. Cases (a) and (b), respectively, represent antiferromagnets with compensated and fully uncompensated interfaces with S. Case (c) depicts a ferrimagnet with a compensated interface. In this case, the symmetry of interfacial coupling between S and the two sublattices is broken~\cite{Gepraegs2016,Cramer2017} by, for example, different wavefunction clouds associated with the localized moments that comprise the sublattice. Interfacial disorder, accounted for in our model, is not depicted explicitly here.}
			\label{fig:schematic}
		\end{center}
	\end{figure*}
	
	In this Letter, we suggest employing insulating AFMs, with their uncompensated surfaces, to induce an effective exchange field in an adjacent superconducting layer. To the best of our knowledge, only FMs have been employed to this end so far. AFMs offer several advantages over FMs in this regard~\cite{Baltz2016,Gomonay2014,Jungwirth2016}. These include minimization of stray magnetic fields, the possibility of electrical tunability~\cite{He2010,Wadley2016,Kosub2017}, avoiding parasitic negative effects of low-energy magnon excitations~\cite{Keffer1952,Kamra2017B} and so on. The proximity effect due to metallic antiferromagnets has been investigated experimentally \cite{Hubener2002} and theoretically~\cite{Moor2012}. Antiferromagnetically ordered impurity chains may also give rise to Majorana state~\cite{Heimes2014}.
	
	Considering a two-sublattice magnetic insulator (MI)/superconductor (S) bilayer structure, we derive the Gor'kov equation for the matrix Green's function in S from a microscopic Hamiltonian including the interface~\cite{Kopnin2001}. Our model for MI encompasses the full range of single-domain magnets from ferro- to antiferro- via ferrimagnets~\cite{Kamra2017B,Kamra2017}. We explicitly include interfacial disorder in our model and find that the induced exchange field is resistant to it, within the Born approximation. We find that the effect of the MI layer is captured by a self-energy which includes interfacial disorder-mediated terms, in addition to the spin splitting term. The latter is found to be large for an uncompensated interface with an AFM. For the system considered here, with the Hamiltonian diagonal in spin space~\footnote{Contributions, non-diagonal in spin space, to the Hamiltonian may arise owing to different physical origins. These include inhomogeneous magnetization, coupling to multiple magnets with non-collinear magnetizations, and unconventional pairing between same-spin electrons in the superconductor.}, the interfacial disorder-mediated terms take a form identical to spin-independent impurity and spin-flip scattering. A third disorder contribution breaks the particle-hole and spin symmetries, but predominantly renormalizes the normal state properties leaving the superconducting state essentially unaffected.

	
	{\it Model and Hamiltonian.} $-$ We consider a MI/S bilayer (Fig. \ref{fig:schematic}) with the S thickness $d_\mathrm{S}$ much smaller than the superconducting coherence length. MI is comprised by a single-domain two-sublattice magnetic insulator where sublattice magnetizations are considered static and collinear to the z-axis. We consider S to be a BCS superconductor in the weak coupling regime such that the Hamiltonian in the grand canonical ensemble reads~\cite{Kopnin2001}:
	\begin{align}\label{eq:hamil}
		\tilde{\mathcal{H}} = & \int d^3r \left[ \sum_{\alpha} \tilde{\psi}_{\alpha}^\dagger (\pmb{r}) \left[ - \partial^2 + V_s (\pmb{r}) \left( \delta_{\alpha \uparrow} - \delta_{\alpha \downarrow} \right) \right] \tilde{\psi}_{\alpha} (\pmb{r}) \right. \nonumber \\
		&  \left. + \sum_{\alpha,\beta} \frac{g}{2} \tilde{\psi}_{\beta}^\dagger (\pmb{r}) \tilde{\psi}_{\alpha}^\dagger (\pmb{r}) \tilde{\psi}_{\alpha} (\pmb{r}) \tilde{\psi}_{\beta} (\pmb{r})  \right].
	\end{align}
	Here, $\tilde{\psi}_{\alpha} (\pmb{r})$ is the electron annihilation operator for z-projected spin $\alpha$ at position $\pmb{r}$,  $\partial^2 \equiv  \nabla^2/2m + \mu - V_i (\pmb{r})$, $\mu$ is the chemical potential, $m$ is electron effective mass, $V_{i}(\pmb{r}) [V_{s}(\pmb{r})]$ represents the spin-independent (dependent) potential energy, $g~(< 0)$ parametrizes the electron-electron attraction, and we have set $\hbar$ to 1. All operators are in the Heisenberg picture and are decorated by a $\tilde{ }$ above. The interface with MI results in the potential energy terms $V_{i,s} (\pmb{r})$. For simplicity, we do not explicitly include bulk contributions to the potential energy here. 
	
	The MI/conductor interface is typically modeled as an effective exchange interaction between the spin densities on the two sides~\cite{Kamra2017}:
	\begin{align}\label{eq:hint}
		\tilde{\mathcal{H}}_{\mathrm{int}} & =  - \int d^2 s \sum_{\mathrm{\mathcal{G} = A,B}} \left[  J_{\mathrm{\mathcal{G}}} \tilde{\pmb{S}}_{\mathrm{\mathcal{G}}}(\pmb{s}) \cdot \tilde{\pmb{S}}(\pmb{s})  \right].
	\end{align}
	Here, $\pmb{s}$ is the two-dimensional position vector in the interfacial plane defined by $y = 0$, $\tilde{\pmb{S}}$ is the electronic spin density operator in S, and $\tilde{\pmb{S}}_{\mathrm{A(B)}}$ is MI sublattice A (B) spin density operator. $J_{\mathrm{A(B)}}$ parametrizes the exchange strength between the MI sublattice A (B) and the S electrons, and depends upon the details of the interface such as its microstructure (Fig. \ref{fig:schematic}). The magnetic spin densities are related to the corresponding magnetizations via the sublattice gyromagnetic ratios $\gamma_{\mathrm{A,B}}$, assumed negative, $\tilde{\pmb{M}}_{\mathrm{A,B}} = - |\gamma_{\mathrm{A,B}}|\tilde{\pmb{S}}_{\mathrm{A,B}}$. We consider sublattice A (B) to be saturated along positive (negative) z-direction with saturation magnetization $M_{\mathrm{A0 (B0)}}$. 
	
	Augmenting the interfacial interaction above [Eq. (\ref{eq:hint})] with a spin-independent contribution and disorder, the net interfacial Hamiltonian may be expressed as:
	\begin{align}\label{eq:hintf}
		\tilde{\mathcal{H}}_{\mathrm{int}}  = &  \int d^3r  \sum_{\alpha} \tilde{\psi}_{\alpha}^\dagger (\pmb{r})  U(\pmb{s}) \delta(y) \left[ a + b  \left( \delta_{\alpha \uparrow} - \delta_{\alpha \downarrow} \right) \right] \tilde{\psi}_{\alpha} (\pmb{r}) ,
	\end{align}
	where $a$ parametrizes the spin-independent contribution of the interfacial interaction and $b = J_{\mathrm{A}} M_{\mathrm{A0}} / 2 |\gamma_{\mathrm{A}}| - J_{\mathrm{B}} M_{\mathrm{B0}} / 2 |\gamma_{\mathrm{B}}|$. $U(\pmb{s})$ accounts for the interfacial disorder which is modeled in a manner analogous to the treatment of impurities-mediated disorder in a bulk conductor~\cite{Kopnin2001,Mahan2000}:
	\begin{align}\label{eq:dispot}
		U(\pmb{s}) = 1 + \sum_{\pmb{s}_i} u(\pmb{s} - \pmb{s}_i),
	\end{align}
	with $u(\pmb{s}- \pmb{s}_i)$ representing the fluctuation in potential energy associated with a `disorder center' located at $\pmb{s}_i$, and we assume $\int d^2 s \ u(\pmb{s}) = 0$. Employing Eq. (\ref{eq:hintf}), the potential energy contribution to the total Hamiltonian [Eq. (\ref{eq:hamil})] corresponds to $V_{i[s]}(\pmb{r}) = U(\pmb{s}) \delta(y) a[b] $.

	
	{\it Gor'kov equation.} $-$ We now formulate the problem at hand in terms of imaginary-time Green's functions in Nambu-spin space. Decorating four-dimensional entities (vectors and matrices) by a $\check{ }$ and two-dimensional by a $\hat{ }$ above, we define $\check{\Psi}^\dagger \equiv \left[ \tilde{\psi}_{\uparrow}^\dagger , \tilde{\psi}_{\downarrow}^\dagger , \tilde{\psi}_{\downarrow} , \tilde{\psi}_{\uparrow} \right] $. We further define the matrix, imaginary-time Green's function as~\cite{Kopnin2001}:
	\begin{align}
		\check{G}(x_1,x_2) \equiv & - \hat{\tau}_z \otimes \hat{\sigma}_0  \mats{ \check{\Psi}(x_1) \check{\Psi}^\dagger(x_2)}, \nonumber \\
		& = \begin{bmatrix}
			G_{\uparrow \uparrow} & G_{\uparrow \downarrow} & F_{\uparrow \downarrow} & F_{\uparrow \uparrow} \\
			G_{\downarrow \uparrow} & G_{\downarrow \downarrow} & F_{\downarrow \downarrow} & F_{\downarrow \uparrow} \\
			- \bar{F}_{\downarrow \uparrow} & - \bar{F}_{\downarrow \downarrow} &  \bar{G}_{\downarrow \downarrow} &  \bar{G}_{\downarrow \uparrow} \\
			- \bar{F}_{\uparrow \uparrow} & - \bar{F}_{\uparrow \downarrow} &  \bar{G}_{\uparrow \downarrow} &  \bar{G}_{\uparrow \uparrow}
		\end{bmatrix},
	\end{align}
	where $\tau = i t$ is the imaginary time, $x_1 \equiv (\pmb{r}_1,\tau_1)$, $\hat{\tau}_{0,x,y,z}$ and $\hat{\sigma}_{0,x,y,z}$ are the identity and Pauli matrices in, respectively, Nambu and spin spaces, and the outer product is expanded as:
	\begin{align}
		\hat{\tau}_z \otimes \hat{\sigma}_0 = & \begin{bmatrix}
			\hat{\sigma}_0 & 0 \\
			0 & - \hat{\sigma}_0
		\end{bmatrix}. \nonumber
	\end{align}
	
	Employing Heisenberg equation of motion for $\tilde{\psi}_{\alpha}(x_1)$ with the Hamiltonian given by Eq. (\ref{eq:hamil}), we obtain the dynamical equation for $G_{\alpha \beta}(x_1,x_2)$:
	\begin{widetext}
		\begin{align}
			\frac{\partial G_{\alpha \beta} (x_1,x_2)}{\partial \tau_1} = & - \delta_{\alpha \beta} \delta (x_1 - x_2) + \left[ \partial_1^2 - V_s(\pmb{r}_1) \left( \delta_{\alpha \uparrow} - \delta_{\alpha \downarrow} \right)  \right] G_{\alpha \beta} (x_1,x_2) - i \sum_{\gamma} \Delta_{\alpha \gamma}(x_1) \bar{F}_{\gamma \beta} (x_1,x_2) ,
		\end{align}
	\end{widetext}
	where $\Delta_{\alpha \beta} (x) \equiv i |g| F_{\alpha \beta} (x,x) $. In simplifying the four-point correlator above, we have employed Wick's theorem~\footnote{Strictly speaking, Wick's theorem is not valid here since our Hamiltonian contains terms fourth order in the ladder operators. However, application of Wick's theorem here is equivalent to the mean-field approximation made in the BCS theory. Please refer to the discussion by Kopnin for further details~\cite{Kopnin2001}.} and disregarded terms which lead to a mere renormalization of the chemical potential~\cite{Kopnin2001}. Dynamical equations for the other components of the matrix Green's function can be derived in an analogous manner~\cite{Kopnin2001}. All these equations may be expressed as a single Gor'kov equation for the matrix Green's function:
	\begin{align}
		\check{\mathcal{G}}^{-1}(x_1) \check{G}(x_1,x_2) = \delta(x_1 - x_2) \hat{\tau}_0 \otimes \hat{\sigma}_0,
	\end{align}
	where
	\begin{align}\label{eq:ginv}
		\check{\mathcal{G}}^{-1}(x_1) = & - \frac{\partial}{\partial \tau_1} \hat{\tau}_z \otimes \hat{\sigma}_0 + \partial_1^2 \hat{\tau}_0 \otimes \hat{\sigma}_0 \nonumber \\
		& - V_s(\pmb{r}_1) \hat{\tau}_z \otimes \hat{\sigma}_z - \check{\Delta}(\pmb{r}_1).
	\end{align}
	For a homogeneous superconducting state, the pair potential matrix may be chosen as $\check{\Delta}(\pmb{r}) =  -i \Delta \hat{\tau}_y \otimes \hat{\sigma}_z$ with real $\Delta$~\cite{Belzig1999,Kopnin2001}.

	
	{\it Interfacial self energy.} $-$ Since the Gor'kov equation can rarely be solved exactly, we resort to perturbation theory within the Green's function method~\cite{Mahan2000} and obtain the self energy arising from the interfacial contribution to the Hamiltonian [Eq. (\ref{eq:hintf})]. To this end, we express $\check{\mathcal{G}}^{-1}(x_1) = \check{\mathcal{G}}^{-1}_0(x_1) - \check{\mathcal{H}}_{\mathrm{int}}(x_1)$ as the sum of the clean superconducting layer plus the interfacial contribution, which assumes the form [using Eqs. (\ref{eq:hintf}) and (\ref{eq:ginv})]:
	\begin{align}
		\check{\mathcal{H}}_{\mathrm{int}} (x_1) = & ~ U(\pmb{s}_1)  \delta(y_1) \left[ a ~ \hat{\tau}_0 \otimes \hat{\sigma}_{0} + b~ \hat{\tau}_z \otimes \hat{\sigma}_{z}  \right] \nonumber \\
		& \equiv ~ U(\pmb{s}_1) \delta(y_1) ~ \check{t}.
	\end{align} 
	The evaluation of the corresponding self-energy follows the method analogous to the case of impurities-mediated disorder in a bulk conductor~\cite{Mahan2000,Kopnin2001} and is detailed in the Supplemental Material~\cite{SupplMat}. Within this method, the so-called cross-diagram technique~\cite{Mahan2000,Kopnin2001}, the following assumptions are made. (i) The perturbation is assumed small thus making the Born approximation. (ii) We average over the positions $\pmb{s}_i$ of the disorder centers. (iii) All diagrams with intersecting impurity scattering lines may be disregarded. (iv) We further neglect diagrams with more than two scattering events. In addition, we employ the quasi-classical approximation in treating the homogeneous superconducting state. With these assumptions, diagrams of all orders can be summed~\cite{Kopnin2001,Mahan2000} and we obtain the main result of this Letter:
	\begin{widetext}
		\begin{align}\label{eq:main}
			\check{\Sigma}_{\mathrm{int}} (\omega_n,\pmb{p}) = & ~ \frac{1}{d_S} \left[ \check{t}  + N_{\mathrm{dis}} \int \frac{d^3 p_1}{(2 \pi)^3} \ |u(\pmb{\kappa} - \pmb{\kappa}_1)|^2 \ \check{t} \ \check{G} (\omega_n,\pmb{p}_1) \ \check{t}  \right],
		\end{align}
	\end{widetext}
	where the result is expressed concisely in the frequency $\omega_n$ and momentum $\pmb{p}$ representation~\cite{SupplMat}. Here, $u(\pmb{\kappa}) \equiv \int d^2 s \ u(\pmb{s}) \ \exp (- i \pmb{\kappa} \cdot \pmb{s})$, $N_{\mathrm{dis}}$ is the areal density of disorder centers, $\pmb{\kappa}$ is the in-plane component of the momentum $\pmb{p}$, and $d_S$ is the thickness of S layer assumed to be much smaller than the superconducting coherence length. The Green's function for the proximity-coupled superconducting layer is given by $\check{G}^{-1}(\omega_n,\pmb{p}) = \check{G}_{0}^{-1}(\omega_n,\pmb{p}) - \check{\Sigma}_{\mathrm{int}} (\omega_n,\pmb{p})$, in terms of the unperturbed Green's function $\check{G}_{0}(\omega_n,\pmb{p})$ and the self-energy evaluated above.

	
	{\it Discussion.} $-$ The self energy [Eq. (\ref{eq:main})], stemming from the interface with MI, comprises a contribution independent of, and thus resistant to, interfacial disorder and a term proportional to the areal density of disorder centers $N_{\mathrm{dis}}$. Apart from a small  renormalization of the chemical potential, the former contribution is simply the effective exchange field, $\propto b = J_{\mathrm{A}} M_{\mathrm{A0}} / 2 |\gamma_{\mathrm{A}}| - J_{\mathrm{B}} M_{\mathrm{B0}} / 2 |\gamma_{\mathrm{B}}|$, induced in S. Thus, an AFM with uncompensated surface, for which $J_{\mathrm{A}} \neq J_{\mathrm{B}}$, $M_{\mathrm{A0}} = M_{\mathrm{B0}}$, and $\gamma_{\mathrm{A}} = \gamma_{\mathrm{B}}$, induces spin splitting in the adjacent S layer.
	
	The interfacial disorder-mediated contribution to the self energy can be further divided into three terms with the integrands in Eq. (\ref{eq:main}) respectively proportional to (i) $a^2 \left( \hat{\tau}_0 \otimes \hat{\sigma}_{0} \check{G} \hat{\tau}_0 \otimes \hat{\sigma}_{0} \right)$, (ii) $ b^2 \left( \hat{\tau}_z \otimes \hat{\sigma}_{z} \check{G} \hat{\tau}_z \otimes \hat{\sigma}_{z} \right)$, and (iii) $ab \left( \hat{\tau}_0 \otimes \hat{\sigma}_{0} \check{G} \hat{\tau}_z \otimes \hat{\sigma}_{z} + \hat{\tau}_z \otimes \hat{\sigma}_{z} \check{G} \hat{\tau}_0 \otimes \hat{\sigma}_{0} \right)$. The term (i) looks like the self energy due to non-magnetic impurities~\cite{Kopnin2001}. Assuming isotropic scattering, this contribution drops out of the superconducting gap as well as the Eilenberger equations for s-wave superconductors, in consistence with the Anderson theorem~\cite{Anderson1959}. Assuming that $\check{\mathcal{G}}^{-1}(x_1)$ is diagonal in spin space, which is the case here [Eq. (\ref{eq:ginv})], the total matrix Green's function is also diagonal in spin space. Taking this into consideration, term (ii) may be rewritten as $ \propto \hat{\tau}_z \otimes \hat{\sigma}_{0} \check{G} \hat{\tau}_z \otimes \hat{\sigma}_{0}$, which has the same form as the self energy contribution due to spin-flip scattering via magnetic impurities~\cite{Kopnin2001}. The effect of such a term has been studied and is known to result in phenomena such as gapless superconductivity~\cite{Maki1969}. It also has consequences for the density of states \cite{Cottet2009,Ouassou2017,Belzig2017} and leads to an enhancement of the Seebeck effect in magnet/superconductor heterostructures~\cite{Rezaei2017}. 
	
	Again, accounting for the diagonal in spin space structure of the total Green's function, the contribution to the self-energy corresponding to the term (iii) assumes the matrix structure $\propto \hat{\tau}_0 \otimes \hat{\sigma}_{z}$, thereby breaking the symmetries in both Nambu and spin spaces. An explicit evaluation of the quasi-classical Green's function matrix shows that this term drops out on integrating over the excitation energy. Thus, this term renormalizes the normal-state properties of the S layer while dropping out in the quasi-classical description of the superconducting state. The analogous term in the self-energy evaluated beyond the Born approximation for magnetic impurities in a bulk superconductor, which does not lead to any spin splitting, was found to break the particle-hole symmetry~\cite{Kalenkov2012}. Its key manifestation was asymmetric scattering with Yu-Shiba-Rusinov states~\cite{Yu1965,Shiba1968,Rusinov1969} resulting in a large thermoelectric effect~\cite{Kalenkov2012}. 
	
	In general, the Hamiltonian, and thus the total matrix Green's function, may be non-diagonal in spin space when, for example, the magnetization is spatially inhomogeneous or the superconductor exhibits unconventional same-spin electron pairing. Under those circumstances, terms (ii) and (iii) may not be interpreted as discussed above.  
	
	Here, we have considered a superconducting layer much thinner than the coherence length. For a thick superconductor, the evaluated self-energy may be incorporated in the boundary conditions for the Gor'kov equation in the bulk. Thus, our theory also provides a microscopic derivation of the boundary conditions describing the interface of a superconductor with a magnetic insulator, complementary to the corresponding evaluations within a scattering theory approach~\cite{Tokuyasu1988,Huertas2002,Eschrig2015a}. Furthermore, we have considered a single-domain magnet leaving possible generalizations to textured and multi-domain interfaces for future work~\cite{Malozemoff1987}. Reference~\cite{Manna2014} reviews exchange bias and magnetic proximity effect together thereby delineating the connection between the two phenomena further and providing directions for generalizing our results.
	
	From the experimental point of view, it is considered difficult to grow metals on insulators due to lattice mismatch. Such interfaces are inevitably disordered. Nevertheless, a strong interfacial exchange coupling has been observed in a wide range of such structures~\cite{Weiler2013,Heinrich2011,Kamra2014,Chumak2015,Uchida2010,Czeschka2011,Gepraegs2016,Cramer2017}. This is consistent with our result which demonstrates that interfacial disorder does not lead to any qualitative changes in physics and the induced exchange field is resistant to this disorder. It, however, leads to additional spin-flip scattering like contributions which, in some cases~\cite{Linder2007,Rezaei2017,Kalenkov2012,Dutta2017}, may be desirable.
	
	As elaborated in the supplemental material~\cite{SupplMat}, the existing literature on exchange bias~\cite{Nogues1999} and spin-mixing conductance~\cite{Brataas2000,Kajiwara2010,Czeschka2011,Weiler2013} provides valuable guidance regarding materials and corresponding expected spin splittings. Several AFMs, such as CoO, Fe$\mathrm{F}_2$, and FeS, may induce fields greater than 100 mT in a 10 nm thick superconducting layer~\cite{SupplMat,Nogues1999}. Furthermore, multilayers incorporating one or more ferromagnetic seed layers are expected to be particularly effective~\cite{SupplMat,Nogues1999}, while still circumventing the disadvantages of spin splitting induced via a ferromagnetic layer. 
	
	
	{\it Summary.} $-$ We have derived and solved the Gor'kov equation for two-sublattice magnetic insulator/thin superconductor bilayer structures. Starting with a microscopic description of the interface, we have evaluated the interfacial self-energy for the matrix (Nambu-spin space) Green's function in the superconducting layer. Our findings show that an antiferromagnet with an uncompensated surface, in addition to ferrimagnets, induces interfacial disorder-resistant spin splitting in the adjacent superconductor. Additional contributions mimicking non-magnetic impurities and spin-flip scattering result due to the interfacial disorder. Our findings, in conjunction with related experiments~\cite{Czeschka2011,Nogues1999,He2010,Kosub2017}, pave the way for employing antiferromagnetic insulators in inducing exchange field in an adjacent superconductor, thereby addressing the feasibility of a wide range of concepts and devices involving spin-split superconductors.

	{\it Acknowledgments.} We thank Juan Carlos Cuevas and Jabir Ali Ouassou for valuable discussions. We acknowledge financial support from the Alexander von Humboldt Foundation, the DFG through SFB 767 and SPP 1538 ``SpinCaT'', and the Research Council of Norway through its Centers of Excellence funding scheme, project 262633, ``QuSpin''.

	\bibliography{proximity}


\widetext
\clearpage
\setcounter{equation}{0}
\setcounter{figure}{0}
\setcounter{table}{0}
\makeatletter
\renewcommand{\theequation}{S\arabic{equation}}

\begin{center}
	\textbf{\large Supplementary material with the manuscript Spin splitting induced in a superconductor by an antiferromagnetic insulator by} \\
	\vspace{0.3cm}
	Akashdeep Kamra, Ali Rezaei, and Wolfgang Belzig
	\vspace{0.2cm}
\end{center}

\setcounter{page}{1}


\section{Frequency-momentum representation}
Adapting Mahan's convention and assuming a time-invariant system~\cite{Mahan2000}, we express a function $f(x_1,x_2)$ in Matsubara frequency representation as follows:
\begin{align}
	f(x_1,x_2) \equiv f(\tau,\pmb{r}_1,\pmb{r}_2) = & \frac{1}{\beta} \sum_{n} e^{- i \omega_n \tau} F(\omega_n,\pmb{r}_1,\pmb{r}_2), \\
	F(\omega_n,\pmb{r}_1,\pmb{r}_2) = & \int_{0}^{\beta} e^{i \omega_n \tau} f(\tau,\pmb{r}_1,\pmb{r}_2) ~ d\tau,
\end{align}
where $\beta \equiv 1/k_B T$ with $k_B$ the Boltzmann constant and $T$ the temperature, $x_1 \equiv \{ \pmb{r}_1, \tau_1 \}$ and so on, $\tau \equiv \tau_1 - \tau_2$, $\omega_n \equiv (2n + 1) \pi / \beta$ are the Matsubara frequencies for a Fermionic system. In the following, we drop the explicit distinction between the functions $f()$ and $F()$. The function being referred to is deemed understood based on its arguments. For example, $f(\omega_n,\pmb{r}_1,\pmb{r}_2)$ from this point on represents what we have called $F(\omega_n,\pmb{r}_1,\pmb{r}_2)$ above. Furthermore, the argument $\omega_n$ is assumed to be implicit in the following discussion.

A general function $f(\pmb{r}_1,\pmb{r}_2)$ is expressed in the momentum representation:
\begin{align}
	f(\pmb{r}_1,\pmb{r}_2) = & \int \frac{d^3 k_1}{(2\pi)^3}  \frac{d^3 k_2}{(2\pi)^3} ~ F(\pmb{k}_1,\pmb{k}_2) ~ e^{i \pmb{k}_1 \cdot \pmb{r}_1} e^{i \pmb{k}_2 \cdot \pmb{r}_2}, \\
	f\left(\pmb{r}^\prime + \frac{\pmb{r}}{2},\pmb{r}^\prime - \frac{\pmb{r}}{2} \right) = & \int \frac{d^3 p}{(2\pi)^3}  \frac{d^3 k}{(2\pi)^3} ~ F\left(\pmb{p} + \frac{\pmb{k}}{2},-\pmb{p} + \frac{\pmb{k}}{2}\right) ~ e^{i \pmb{p} \cdot \pmb{r}} e^{i \pmb{k} \cdot \pmb{r}^\prime},
\end{align}
where $\pmb{p} \equiv (\pmb{k}_1 - \pmb{k}_2)/2$, $\pmb{k} \equiv \pmb{k}_1 + \pmb{k}_2$, $\pmb{r} \equiv \pmb{r}_1 - \pmb{r}_2$, and $\pmb{r}^\prime \equiv (\pmb{r}_1 + \pmb{r}_2)/2$. With the definitions:
\begin{align}
	f_1\left(\pmb{r},\pmb{r}^\prime\right) \equiv f\left(\pmb{r}^\prime + \frac{\pmb{r}}{2},\pmb{r}^\prime - \frac{\pmb{r}}{2} \right),\\
	F_1\left(\pmb{p},\pmb{k}\right) \equiv F\left(\pmb{p} + \frac{\pmb{k}}{2},-\pmb{p} + \frac{\pmb{k}}{2}\right),
\end{align}
we can describe the function in the relative and center of mass coordinates representation:
\begin{align}
	f_1\left(\pmb{r},\pmb{r}^\prime\right) = & \int \frac{d^3 p}{(2\pi)^3}  \frac{d^3 k}{(2\pi)^3} ~ F_1\left(\pmb{p},\pmb{k}\right) ~ e^{i \pmb{p} \cdot \pmb{r}} e^{i \pmb{k} \cdot \pmb{r}^\prime}, \\
	F_1\left(\pmb{p},\pmb{k}\right) = & \int d^3 \pmb{r} d^3 \pmb{r}^\prime ~ f_1\left(\pmb{r},\pmb{r}^\prime\right) ~ e^{-i \pmb{p} \cdot \pmb{r}} e^{-i \pmb{k} \cdot \pmb{r}^\prime}.
\end{align}
This representation allows us to treat the variations in functions on small (inverse Fermi momentum) and large (superconducting coherence) length scales effectively~\cite{Kopnin2001}. In particular, the description of a spatially homogeneous system can be treated as independent of $\pmb{r}^\prime$ and may be developed in terms of $\pmb{p}$ alone, while disregarding $\pmb{k}$. Once again, in the following, and in the main text, we do not explicitly distinguish between the different functions (e.g. $f(),~f_1(), ~F()$ and so on). We employ the same letters to represent the function appropriate for that particular representation, which, in turn, becomes evident from the arguments specifying the function. For example, $f(\pmb{p},\pmb{k})$ is understood to represent the following expression in terms of the real-space function $f(x_1,x_2) \equiv f(\tau,\pmb{r}_1,\pmb{r}_2)$:
\begin{align}
	f\left(\pmb{p},\pmb{k}\right) = &  \int_{0}^{\beta} d\tau \int d^3 \pmb{r} d^3 \pmb{r}^\prime ~ f\left(\tau,\pmb{r}^\prime + \frac{\pmb{r}}{2},\pmb{r}^\prime - \frac{\pmb{r}}{2} \right) ~ e^{i \omega_n \tau} e^{-i \pmb{p} \cdot \pmb{r}} e^{-i \pmb{k} \cdot \pmb{r}^\prime}.
\end{align}


\section{Perturbative evaluation of Green's function}
Expressing the problem to be solved as the sum of an unperturbed and a perturbation contributions $\check{\mathcal{G}}^{-1}(x_1) = \check{\mathcal{G}}^{-1}_0(x_1) - \check{\mathcal{H}}_{\mathrm{int}}(x_1)$, the total Green's function matrix can be expanded as a sum of contributions to increasing degrees in the perturbation:
\begin{align}
	\check{G}\left(x_1,x_2\right) = & \sum_{n = 0}^{\infty} \check{G}^{(n)}\left(x_1,x_2\right),
\end{align}
where $\check{G}^{(0)}\left(x_1,x_2\right)$ is the Green's function matrix for the unperturbed problem. For the case at hand, this corresponds to a superconducting film without the magnet.  Substituting above form of the Green's function into the Gor'kov equation and switching to frequency representation, we obtain the following recursive relations~\cite{Mahan2000,Kopnin2001}:
\begin{align}
	\check{G}^{(n)}\left(\pmb{r}_1,\pmb{r}_2\right) = & \int d^3 r_3~ \check{G}^{(0)}\left(\pmb{r}_1,\pmb{r}_3\right) \check{\mathcal{H}}_{\mathrm{int}}(\pmb{r}_3) \check{G}^{(n-1)}\left(\pmb{r}_3,\pmb{r}_2\right),
\end{align}
with $n \geq 1$. Since we work within the quasi-classical approximation for superconductivity assuming the unperturbed solution to represent a homogeneous superconducting state, the corresponding Green's function matrix can be represented as:
\begin{align}\label{g0}
	\check{G}^{(0)}\left(\pmb{r}_1,\pmb{r}_2\right) = & \int \frac{d^3 p}{(2\pi)^3} \check{G}^{(0)}\left(\pmb{p}\right) e^{i \pmb{p} \cdot (\pmb{r}_1 - \pmb{r}_2)}.
\end{align}
This representation will be used repeatedly in the following analysis.

The first order correction can be simplified to:
\begin{align}\label{g11}
	\check{G}^{(1)}\left(\pmb{r}_1,\pmb{r}_2\right) = & \int  ~ \frac{d^3 p_1}{(2\pi)^3} \frac{d^3 p_2}{(2\pi)^3} ~e^{i\left(\pmb{p}_1 \cdot \pmb{r}_1 - \pmb{p}_2 \cdot \pmb{r}_2  \right)} ~ \check{G}^{(0)}\left(\pmb{p}_1\right) \check{\mathcal{H}}_{\mathrm{int}}(\pmb{p}_1 - \pmb{p}_2) \check{G}^{(0)}\left(\pmb{p}_2\right),
\end{align}
where
\begin{align}
	\check{\mathcal{H}}_{\mathrm{int}}(\pmb{p}) \equiv & \int d^3 r ~ \check{\mathcal{H}}_{\mathrm{int}}(\pmb{r}) ~ e^{-i \pmb{p} \cdot \pmb{r}}, \\
	= & \left[ (2 \pi)^2 \delta (\pmb{\kappa}) + u(\pmb{\kappa}) \sum_{\pmb{s}_i}  e^{- i \pmb{\kappa} \cdot \pmb{s}_i} \right] ~ \check{t}, \label{hp}
\end{align} 
with $\pmb{p} \equiv \{\pmb{\kappa}, \xi \}$. In evaluating the above expression, we have employed the conventions and definitions introduced in the main text. Employing Eq. (\ref{hp}) in Eq. (\ref{g11}). the first order correction reduces to:
\begin{align}
	\check{G}^{(1)}\left(\pmb{r}_1,\pmb{r}_2\right) = & \int  ~ \frac{d^3 p_1}{(2\pi)^3} \frac{d^3 p_2}{(2\pi)^3} ~e^{i\left(\pmb{p}_1 \cdot \pmb{r}_1 - \pmb{p}_2 \cdot \pmb{r}_2  \right)} ~ \check{G}^{(0)}\left(\pmb{p}_1\right)  (2 \pi)^2 \delta (\pmb{\kappa}_1 - \pmb{\kappa}_2) ~ \check{t} ~  \check{G}^{(0)}\left(\pmb{p}_2\right) \quad + \nonumber \\
	&   \int  ~ \frac{d^3 p_1}{(2\pi)^3} \frac{d^3 p_2}{(2\pi)^3} ~e^{i\left(\pmb{p}_1 \cdot \pmb{r}_1 - \pmb{p}_2 \cdot \pmb{r}_2  \right)} ~ \sum_{\pmb{s}_i}  e^{- i (\pmb{\kappa}_1 - \pmb{\kappa}_2) \cdot \pmb{s}_i} ~\check{G}^{(0)}\left(\pmb{p}_1\right)   u (\pmb{\kappa}_1 - \pmb{\kappa}_2) ~ \check{t} ~  \check{G}^{(0)}\left(\pmb{p}_2\right), \\
	= & \int  ~ \frac{d^3 p_1}{(2\pi)^3} \frac{d \xi_2}{2\pi} ~ e^{i \left(\xi_1 y_1 - \xi_2 y_2 \right)} ~ e^{i \pmb{\kappa}_1 \cdot \left( \pmb{s}_1 - \pmb{s}_2 \right)} ~ \check{G}^{(0)}\left(\pmb{p}_1\right) ~ \check{t} ~  \check{G}^{(0)}\left(\pmb{\kappa}_1,\xi_2 \right) \quad + \nonumber \\
	&   N_{\mathrm{dis}} \int  ~ \frac{d^3 p_1}{(2\pi)^3} \frac{d \xi_2}{2\pi} ~ e^{i \left(\xi_1 y_1 - \xi_2 y_2 \right)} ~ e^{i \pmb{\kappa}_1 \cdot \left( \pmb{s}_1 - \pmb{s}_2 \right)} ~ u(0) ~\check{G}^{(0)}\left(\pmb{p}_1\right) ~ \check{t} ~  \check{G}^{(0)}\left(\pmb{\kappa}_1,\xi_2 \right),
\end{align}
where we have averaged over the disorder center locations via the replacement $\sum_{\pmb{s}_i} \to N_{\mathrm{dis}} \int d^2 s_i$, and $u(0) = \int d^2s ~ u(s) = 0$ results in the second term vanishing:
\begin{align}
	\check{G}^{(1)}\left(\pmb{r}_1,\pmb{r}_2\right) = & \int  ~ \frac{d^3 p_1}{(2\pi)^3} \frac{d \xi_2}{2\pi} ~ e^{i \left(\xi_1 y_1 - \xi_2 y_2 \right)} ~ e^{i \pmb{\kappa}_1 \cdot \left( \pmb{s}_1 - \pmb{s}_2 \right)} ~ \check{G}^{(0)}\left(\pmb{p}_1\right) ~ \check{t} ~  \check{G}^{(0)}\left(\pmb{\kappa}_1,\xi_2 \right).
\end{align}
The expression obtained above describes an inhomogeneous system due to the breaking of translational invariance by the interface. However, within the quasi-classical approximation, we expect a homogeneous superconducting state. Thus the expression above goes beyond the quasi-classical limit. We obtain the contribution relevant for describing superconductivity, stemming from a narrow phase-space around the Fermi energy, by averaging over the thickness $d_S$ of the superconductor:
\begin{align}
	\left \langle \check{G}^{(1)}\left(\pmb{r}_1,\pmb{r}_2\right) \right \rangle = & \frac{1}{d_S} \int dy^\prime~ \check{G}^{(1)}\left(\pmb{r}_1,\pmb{r}_2\right), \\
	= & \frac{1}{d_S}  \int  ~ \frac{d^3 p_1}{(2\pi)^3} \frac{d \xi_2}{2\pi} dy^\prime ~ e^{i y^\prime \left(\xi_1 - \xi_2 \right)} ~ e^{i y \frac{\left(\xi_1 + \xi_2 \right)}{2}} ~ e^{i \pmb{\kappa}_1 \cdot \left( \pmb{s}_1 - \pmb{s}_2 \right)} ~ \check{G}^{(0)}\left(\pmb{p}_1\right) ~ \check{t} ~  \check{G}^{(0)}\left(\pmb{\kappa}_1,\xi_2 \right),
\end{align}
where $y^\prime = (y_1 + y_2)/2$ and $y = y_1 - y_2$. This leads us to our result for the first order correction:
\begin{align}
	\left \langle \check{G}^{(1)}\left(\pmb{r}_1,\pmb{r}_2\right) \right \rangle = & \int \frac{d^3 p}{(2\pi)^3} ~ e^{i \pmb{p} \cdot (\pmb{r}_1 - \pmb{r}_2)} ~ \check{G}^{(0)}\left(\pmb{p}\right) ~ \frac{\check{t}}{d_S} ~  \check{G}^{(0)}\left(\pmb{p}\right). 
\end{align}

The evaluation of the second order correction follows an analysis similar to the above. We wish to evaluate $\left \langle \check{G}^{(2)}\left(\pmb{r}_1,\pmb{r}_2\right) \right \rangle$ with
\begin{align}\label{g21}
	\check{G}^{(2)}\left(\pmb{r}_1,\pmb{r}_2\right) = &  \int d^3 r_3~ \check{G}^{(0)}\left(\pmb{r}_1,\pmb{r}_3\right) \check{\mathcal{H}}_{\mathrm{int}}(\pmb{r}_3) \check{G}^{(1)}\left(\pmb{r}_3,\pmb{r}_2\right).
\end{align}
In order to obtain the desired result within our approximation, we make the following replacement:
\begin{align}
	\check{G}^{(1)}\left(\pmb{r}_1,\pmb{r}_2\right) \to &  \left \langle \check{G}^{(1)}\left(\pmb{r}_1,\pmb{r}_2\right) \right \rangle + \nonumber \\
	& \int  ~ \frac{d^3 p_1}{(2\pi)^3} \frac{d^3 p_2}{(2\pi)^3} ~e^{i\left(\pmb{p}_1 \cdot \pmb{r}_1 - \pmb{p}_2 \cdot \pmb{r}_2  \right)} ~ \sum_{\pmb{s}_i}  e^{- i (\pmb{\kappa}_1 - \pmb{\kappa}_2) \cdot \pmb{s}_i} ~ \check{G}^{(0)}\left(\pmb{p}_1\right) ~ u (\pmb{\kappa}_1 - \pmb{\kappa}_2) ~ \check{t} ~ \check{G}^{(0)}\left(\pmb{p}_2\right). \nonumber
\end{align}
The disorder has to be treated separately from the $\left \langle \check{G}^{(1)}\left(\pmb{r}_1,\pmb{r}_2\right) \right \rangle$ term since the pre-averaging procedure fails to capture the disorder-mediated scrambling of momenta. Employing the above replacement, Eq. (\ref{g21}) can be simplified into several contributions. All contributions stemming from scattering by single and multiple  but distinct disorder-centers vanish on account of $u(0) = 0$. The term due to two scattering events from the same disorder-centers leads to a finite result. Combined with the other assumptions of the cross-diagram technique~\cite{Kopnin2001,Mahan2000}, mentioned in the main text, the required second order correction becomes:
\begin{align}
	\left \langle \check{G}^{(2)}\left(\pmb{r}_1,\pmb{r}_2\right) \right \rangle = & \int \frac{d^3 p}{(2\pi)^3} ~ e^{i \pmb{p} \cdot (\pmb{r}_1 - \pmb{r}_2)} ~ \check{G}^{(0)}\left(\pmb{p}\right) ~ \check{\Sigma}^{(1)}_{\mathrm{int}} \left(\pmb{p}\right) ~  \check{G}^{(0)}\left(\pmb{p}\right),
\end{align}
where 
\begin{align}
	\check{\Sigma}^{(1)}_{\mathrm{int}}\left(\pmb{p}\right) = & \frac{N_{\mathrm{dis}}}{d_S} \int \frac{d^3 p_1}{(2\pi)^3} ~ \left|u(\pmb{\kappa} - \pmb{\kappa}_1)\right|^2 ~ \check{t} ~ \check{G}^{(0)}\left(\pmb{p}_1\right) ~ \check{t}.
\end{align}

Proceeding along similar lines employing the approximations introduced above and evaluating the higher order terms, we can sum all terms in a manner analogous to the treatment of bulk impurity scattering within the cross diagram technique~\cite{Kopnin2001,Mahan2000}. The final result is obtained as 
\begin{align}
	\left \langle \check{G}\left(\pmb{r}_1,\pmb{r}_2\right) \right \rangle = & \int \frac{d^3 p}{(2\pi)^3} ~ e^{i \pmb{p} \cdot (\pmb{r}_1 - \pmb{r}_2)} ~ \check{G}\left(\pmb{p}\right),
\end{align}
with the expression for $\check{G}\left(\pmb{p}\right) \equiv \check{G}\left(\omega_n,\pmb{p}\right) $ as given in the main text.


\section{Materials and expected effective fields}
In the present section, we discuss some concrete materials along with the spin splittings, or equivalently effective magnetic fields, expected to be induced by them in an adjacent conductor. We first present these estimates on the basis of the experimental data available from exchange bias studies~\cite{Nogues1999}. The compiled data demonstrates the variation in the induced fields with different materials, textures, and fabrication techniques. In particular, incorporation of a ferromagnetic seed layer can be very effective in achieving a large induced field without necessitating epitaxial growth. Second, we estimate the expected spin splitting from the spin-mixing conductance measured via various experimental studies in ferromagnet/metal bilayers~\cite{Czeschka2011,Weiler2013,Kajiwara2010}. One can expect a perfectly uncompensated interface to induce a comparable field. The resulting estimate is consistent with the highest values expected from the exchange bias data thus suggesting that several systems exhibit nearly perfect uncompensation at the interface. Detailed spin-mixing conductance studies for antiferromagnets, analogous to exchange bias data, are not available at this point.

\subsection{Estimates from exchange bias experiments}
Enlisting a few examples from Ref. \onlinecite{Nogues1999} here, we refer the readers to this review article for a more extensive analysis and further references. The effective fields have been estimated employing the reported interfacial energy densities~\cite{Nogues1999} assuming a nominal magnetization of $2 \times 10^5~\mathrm{A}/\mathrm{m}$ and a $10~\mathrm{nm}$ thick superconductor. The `texture' corresponds to the magnet being polycrystalline (`poly') or one of its specific crystal planes exposed at its interface with the superconductor. `RT' stands for room temperature. A further discussion of the `comments' is also presented below. Some of the discussed materials are not antiferromagnetic at room temperatures, which does not hinder their use with superconductors at low temperatures. 

\begin{table}[hb]
	\begin{center}
		\begin{tabular}{|c|c|c|c|c|}
			\hline
			Material & Temperature (K) & Texture & Effective field (T) & Comments \\
			\hline 
			CoO & 150 & poly & $10^{-1}$ & -- \\
			CoO & 100 & poly-multi & $10^{0}$ & Ferromagnet in mutlilayer \\
			CoO & 77 & (1 1 1) & $10^{-1}$ & -- \\
			FeS & 10 & poly & $10^{-1}$ & -- \\
			Fe$\mathrm{F}_2$ & 10 & (1 1 0) & $10^{0}$ & -- \\
			Fe$\mathrm{F}_2$ & 10 & (0 0 1) & $10^{-3}$ & -- \\
			CrN & 10 & poly & $10^{-1}$ & -- \\
			NiO & RT & poly & $10^{-3} - 10^{-2} $ & -- \\
			NiO & RT & (1 1 1) & $10^{-3} - 10^{-2}$ & Enhancement at low temperatures \\
			NiO & RT & (1 0 0) & $10^{-2} - 10^{-1}$ & -- \\
			$\mathrm{Cr}_2 \mathrm{O}_3$ & RT & poly & $10^{-3}$ & Comparison with recent studies \\ 
			\hline
		\end{tabular}
		\caption{Materials and expected induced effective magnetic fields evaluated via the interfacial energy density data compiled in Ref. \onlinecite{Nogues1999}.}\label{thetable}
	\end{center}
\end{table}

\noindent
{\bf Discussion of comments:}
\begin{itemize}
	\item {\it Ferromagnet in multilayer.} -- Incorporating one or more ferromagnetic layers into a magnetic multilayer terminated with an antiferromagnetic layer allows for a strong sublattice-asymmetry at the exposed surface. Such an arrangement does not depend on having an epitaxial growth. Furthermore, the presence of a ferromagnetic (seed) layer far away from the exposed antiferromagnet surface, at which the superconductor is deposited, does not influence the conductor. The underlying physics is explained by the field cooling effect~\cite{Nogues1999}. The heterostructure is heated above the Neel temperature of the antiferromagnetic layer. It is then allowed to cool gradually in the presence of an applied magnetic field, which keeps the ferromagnet fully aligned. As the antiferromagnet begins to order below its Neel temperature, the atomic layer next to the ferromagnet is fully ordered due to interfacial exchange interaction with the ordered ferromagnet. The subsequent atomic layers in the antiferromagnet follow appropriate ordering, consistent with the antiferromagnetic interaction with the previous atomic layer, resulting in an atomically layered configuration throughout, which gives a strong sublattice symmetry-breaking at the other end of the antiferromagnet, where the conductor is deposited.
	
	\item {\it Enhancement at low temperatures.} -- Reference \cite{Nogues1999} also documents an isolated observation of an effective field larger by three orders of magnitude for the same system at 10 K.
	
	\item {\it Comparison with recent studies.} --  The estimated field here is consistent with the recent experimental observation of exchange bias via epitaxially grown $\mathrm{Cr}_2 \mathrm{O}_3$ with a reasonably rough surface~\cite{He2010}. Their demonstration of electric switching allows for yet another functionality, and is directly applicable to our proposal as well. 
\end{itemize}

\subsection{Estimates from spin-mixing conductance experiments}
Kajiwara and coworkers~\cite{Kajiwara2010} have estimated an exchange coupling energy of 10 meV per `bond' employing the spin transfer studies across yttrium iron garnet (YIG)/platinum interface. This corresponds to a spin-mixing conductance $\sim 10^{19}~ \mathrm{m}^{-2}$. Considering the lattice constant of around 1 nm for YIG, we estimate the interfacial energy density of $\sim 10^{-3} ~ \mathrm{J}/\mathrm{m}^2$. Following an approach analogous to the exchange bias evaluation above, we estimate an effective field of $1 - 10$ T in the adjacent conductor with thickness 10 nm. There is a slight ambiguity in this estimation stemming from a difference in the lattice constants of YIG and a typical conductor. However, since we only estimate the order of magnitude, this ambiguity is of little consequence. Since similar spin-mixing conductances have been measured across a range of magnet/metal bilayers~\cite{Czeschka2011}, we expect the induced fields by the corresponding uncompensated interfaces to be of similar magnitude.

\end{document}